\documentclass[aps,prd,preprint,superscriptaddress,nofootinbib]{revtex4-1}
\usepackage{braket}
\usepackage{xargs}
\usepackage{mathtools}
\usepackage[english]{babel}
\usepackage[utf8]{inputenc}
\usepackage{amsmath,amssymb,braket,slashed,mathrsfs}
\usepackage{pifont}
\usepackage{MnSymbol}
\usepackage{graphicx}
\usepackage{tikz}
\usetikzlibrary{shapes,arrows,positioning}
\tikzset{decision/.style={diamond, draw, fill=blue!20, text width=4.5em, text badly centered, inner sep=0pt}}
\tikzset{block/.style={rectangle, draw, fill=blue!20, text width=10em, text centered, rounded corners, minimum width=3.5cm}}
\tikzset{block1/.style={rectangle, draw, fill=blue!20, text width=18.5em, text centered, rounded corners, minimum width=3.5cm}}
\tikzset{line/.style={draw, -latex, thick}}
\usepackage{color,hyperref}
\usepackage{multirow,array}

\usepackage{cleveref}

\newcommand{\ba}{\begin{eqnarray}}
\newcommand{\ea}{\end{eqnarray}}
\newcommand{\be}{\begin{equation}}
\newcommand{\ee}{\end{equation}}

\newcommand{\nn}{\nonumber}

\newcommand{\innovation}{Collaborative Innovation Center of Quantum Matter, Beijing 100871, China}
\newcommand{\chep}{Center for High Energy Physics, Peking University, Beijing 100871, China}
\newcommand{\pkuphy}{School of Physics and State Key Laboratory of Nuclear Physics and Technology, Peking University, Beijing 100871,
China}

\newcommand{\Uconn}{Department of Physics, University of Connecticut, Storrs, CT 06269, USA}
\newcommand{\RBRC}{RIKEN-BNL Research Center, Brookhaven National Laboratory, Building 510, Upton, NY 11973}
\newcommand{\Mainz}{Helmholtz Institute Mainz, Mainz, Germany}
\newcommand{\GSI}{GSI Helmholtzzentrum f\"ur Schwerionenforschung, Darmstadt, Germany}
\newcommand{\JGU}{Johannes Gutenberg University, Mainz, Germany}
\newcommand{\Bonn}{Helmholtz-Institut f\"ur Strahlen- und Kernphysik and Bethe
Center for Theoretical Physics,\\ Universit\"at Bonn, 53115 Bonn, Germany}

\begin{document}
\title{Lattice QCD calculation of the electroweak box diagrams for the kaon
semileptonic decays}

\author{Peng-Xiang~Ma}\affiliation{\pkuphy}
\author{Xu~Feng}\email{xu.feng@pku.edu.cn}\affiliation{\pkuphy}\affiliation{\innovation}\affiliation{\chep}
\author{Mikhail~Gorchtein}\email{gorshtey@uni-mainz.de}\affiliation{\Mainz}\affiliation{\GSI}\affiliation{\JGU}
\author{Lu-Chang~Jin}\email{ljin.luchang@gmail.com}\affiliation{\Uconn}\affiliation{\RBRC}
\author{Chien-Yeah~Seng}\email{cseng@hiskp.uni-bonn.de}\affiliation{\Bonn}

\date{\today}

\begin{abstract}
We present a lattice QCD calculation of the axial $\gamma
W$-box diagrams relevant for the kaon semileptonic decays.
    We utilize a recently proposed method, which 
    connects the electroweak radiative corrections in Sirlin's representation to that in chiral
    perturbation theory. It allows us to use the axial $\gamma W$-box correction
    in the SU(3) limit to obtain the low energy constants for chiral perturbation
    theory. From first principles our results confirm the previously used low energy constants 
    provided by the minimal resonance model with a significant reduction in uncertainties.

\end{abstract}

\maketitle

\section{Introduction}

In the Standard Model, the Cabibbo-Kobayashi-Maskawa (CKM) matrix is a three-generation quark mixing matrix
which describes how the strength of the flavour-changing weak interaction
in the leptonic sector is distributed among the three quark generations.
The precise determination of the CKM matrix elements is of vital importance in
the stringent test of CKM unitarity and search of new physics beyond the Standard Model.
As quoted in the 2020 Review by the Particle Data Group~\cite{Zyla:2020zbs}, 
there exists a $\sim3$ sigma deviation from unitarity in the first
row of CKM matrix elements
\be
\label{eq:CKM_unitarity}
|V_{ud}|^2+|V_{us}|^2+|V_{ub}|^2=0.9984(3)_{V_{ud}}(4)_{V_{us}}.
\ee
Here $|V_{ub}|^2\approx 1.5\times10^{-5}$ is negligibly small and thus only
$|V_{ud}|$ and $|V_{us}|$ play a role in the unitarity test.

The most precise determination of
$|V_{ud}|=0.97370(10)_{\mathrm{exp}+\mathrm{nucl}}(10)_{\mathrm{RC}}$ quoted in
the 2020 PDG review~\cite{Zyla:2020zbs} stems from
the superallowed nuclear beta decays~\cite{Hardy:2014qxa,Hardy:2020qwl}, with the first uncertainty
arising from the experimental measurements and nuclear physics corrections and
the second one from the electroweak radiative corrections (RCs)\footnote{Notice, however, that this quoted value does not include the contributions from several new nuclear corrections investigated in Refs.\cite{Seng:2018qru,Gorchtein:2018fxl}.}. It is the
update of the RCs from a dispersive analysis~\cite{Seng:2018yzq,Seng:2018qru},
which makes the value of $|V_{ud}|$ about 2$\sigma$ smaller than that in the 2018 PDG
review~\cite{Tanabashi:2018oca}. Very recently, the RCs to the
$\pi_{\ell3}$ decays were calculated using lattice QCD with the focus on the so-called axial $\gamma W$-box
diagrams~\cite{Feng:2020zdc}. 
It allowed for a significant reduction of the hadronic uncertainty in the RCs, and provided an independent cross-check of the dispersion relation analysis of the neutron RCs~\cite{Seng:2020wjq}. 
In the future a direct lattice QCD calculation of the RCs to the neutron decay
could help to further improve the determination of $|V_{ud}|$~\cite{Seng:2019plg}.

The $|V_{us}|$ can be determined from kaon, hyperon or tau
decays, 
with kaon decays providing the best precision.  
Leptonic decays $K\to\mu\nu$ (denoted by $K_{\mu2}$) combined with $\pi\to\mu\nu$ give access to the ratio $|V_{us}/V_{ud}|$, whereas
semileptonic decays $K\to\pi\ell\nu$ (denoted by $K_{\ell3}$) give a handle on $|V_{us}|$ independently.
The traditional way of determining $|V_{us}|$ relies on the experimental
measurements of $K_L^0\to\pi e\nu$ to avoid the isospin-breaking effects
($\pi^0$-$\eta$ mixing) in the charged kaon decays and the complication from the
second (scalar) form factor present
in the muonic decays. Nowadays, due to the high-statistics data collected in the
experiments, the comparison between different decay modes is justified~\cite{Antonelli:2010yf}. The 
decays including $K_L^0\to\pi \ell\nu$, $K^\pm\to\pi^0\ell^\pm \nu$ and
$K_S^0\to\pi e\nu$ with $\ell=e,\mu$ are used to determine $|V_{us}|$ via the
master formula~\cite{Zyla:2020zbs}
\be
\label{eq:master_formula}
\Gamma_{K\ell3}=\frac{G_F^2m_K^5}{192\pi^3}S_{EW}(1+\delta_K^\ell+\delta_{SU2})C^2|V_{us}|^2f_+^2(0)I_K^\ell.
\ee
Here, $\Gamma_{K\ell3}$ is the $K_{\ell3}$ decay width, $G_F$ is the Fermi
constant, $m_K$ is the kaon mass, $S_{EW}$ is the short-distance radiative
correction, $\delta_K^\ell$ is the long-distance radiative correction,
$\delta_{SU2}$ is the strong isospin-violating effect, $C^2$ is 1 for the neutral
kaon decay and $1/2$ for the charged case,
$f_+(q^2)$ is the $K^0\rightarrow \pi^-$ vector form factor and $I_K^\ell$ is the phase-space
integral which contains the information of the momentum dependence in the form factors.
Averaging over the experimental measurements with appropriate theory inputs of various Standard Model corrections, the product $|V_{us}|f_+(0)$ is given as~\cite{Moulson:2017ive}
\be
f_+(0)|V_{us}|=0.2165(4),
\ee
with the uncertainty dominated by the experimental measurements and RCs. The
form factor $f_+(0)$ can be provided by lattice QCD
calculations~\cite{Boyle:2015hfa,Carrasco:2016kpy,Aoki:2017spo,Bazavov:2018kjg,Kakazu:2019ltq}.
The FLAG
average~\cite{Aoki:2019cca} 
for $N_f=2+1+1$ simulations yields $f_+(0)=0.9698(17)$ according to an update on
December 2020, which results in a determination of 
\be
|V_{us}|=0.2232(4)_{\mathrm{exp}+\mathrm{RC}}(4)_{\mathrm{lat}},\quad\mbox{for
}K_{\ell3}.
\ee
High-precision experimental data on $K_{\mu2}$ and $\pi_{\mu2}$
decays~\cite{Marciano:2004uf,Ambrosino:2005fw} also
accurately determine the ratio
$|V_{us}/V_{ud}|f_{K^\pm}/f_{\pi^\pm}=0.2760(4)$~\cite{Moulson:2017ive}.
Employing the FLAG $N_f=2+1+1$ lattice QCD average~\cite{Dowdall:2013rya,Carrasco:2014poa,Bazavov:2017lyh,Miller:2020xhy} for the ratio of decay constants
$f_{K^\pm}/f_{\pi^\pm}=1.1932(21)$, a value of $|V_{us}|=0.2252(5)$ is
obtained, which has a $2.6\sigma$ deviation from the $K_{\ell3}$-based value.
Combining the $|V_{us}|$ from $K_{\ell3}$ and $K_{\mu2}$ decays
yield~\footnote{Here $|V_{us}|$ is slightly different from the PDG value
$0.2245(8)$ due to the update of the FLAG value of $f_+(0)$.
Correspondingly, the value of $|V_{ud}|^2+|V_{us}|^2+|V_{ub}|^2$ given in
Eq.~(\ref{eq:CKM_unitarity}) also slightly differs from the PDG value of
$0.9985(3)_{V_{ud}}(4)_{V_{us}}$.}
\be
|V_{us}|=0.2243(8),\quad\mbox{weighted average of $K_{\ell3}$ and $K_{\mu2}$}.
\ee
It should also be mentioned that $|V_{us}|$ obtained from hyperon and tau decays
are given by $|V_{us}|=0.2250(27)$~\cite{Cabibbo:2003ea} and 0.2221(13)~\cite{Amhis:2019ckw}, respectively, both
having larger uncertainties than the kaon decays.

To gain a better understanding of the violation of the first-row CKM unitarity
in Eq.~(\ref{eq:CKM_unitarity}) and the disagreement in the determination of $|V_{us}|$
between the $K_{\ell3}$ and $K_{\mu2}$, for the $K_{\ell3}$ decays it requires both a more precise
determination of the form factor $f_+(0)$ and a direct calculation of RCs from
lattice QCD. The latter is more challenging due to the inclusion of both weak
and electromagnetic currents in the calculation and is the focus in this paper.

Recently, the horizons of lattice QCD studies have been extended to include various
processes with higher-order electroweak interactions. The examples include kaon
mixing~\cite{Christ:2012se,Bai:2014cva,Christ:2015pwa}, 
rare kaon
decays~\cite{Christ:2015aha,Christ:2016eae,Christ:2016mmq,Bai:2018hqu,Christ:2019dxu,Christ:2020hwe},
double beta
decays\,\cite{Tiburzi:2017iux,Shanahan:2017bgi,Nicholson:2018mwc,Feng:2018pdq,Tuo:2019bue,Detmold:2020jqv,Feng:2020nqj,Davoudi:2020xdv,Davoudi:2020gxs},
inclusive $B$-meson decays\,\cite{Hansen:2017mnd,Hashimoto:2017wqo,Gambino:2020crt}, as well
as the electromagentic and radiative corrections to the weak
decays~\cite{Carrasco:2015xwa,Lubicz:2016xro,Giusti:2017dwk,Feng:2018qpx,Feng:2020mmb,Desiderio:2020oej,Frezzotti:2020bfa}.
Among all these processes, the lattice QCD calculation of RCs in $K_{\ell3}$ still remains one of the
largest challenges as it essentially involves a computation of five-point
correlation functions. In Ref.~\cite{Seng:2020jtz}, it proposes a new method
which bridges
the lattice QCD calculation with chiral perturbation theory
(ChPT)~\cite{Cirigliano:2001mk,Cirigliano:2008wn}. For the $K_{\ell3}$
decay in
the flavor SU(3) limit, it demonstrates that the lattice QCD calculation of the axial $\gamma W$-box
diagrams can provide all unknown low-energy constants (LECs) that enter the long-distance radiative correction $\delta_K^\ell$ in the ChPT representation at the order of
$O(e^2p^2)$,
thus removing the dependence of the RCs on the model used to estimate these LECs.
In this paper we will first briefly introduce the methodology and then
present the lattice calculation of RCs.

\section{Methodology}

We start the discussion of the treatment of RCs in $K_{\ell3}$ decays with two
theoretical frameworks:
Sirlin's representation and
the ChPT representation.

     \begin{figure}[htb]
     \centering
         \includegraphics[width=0.5\textwidth,angle=0]{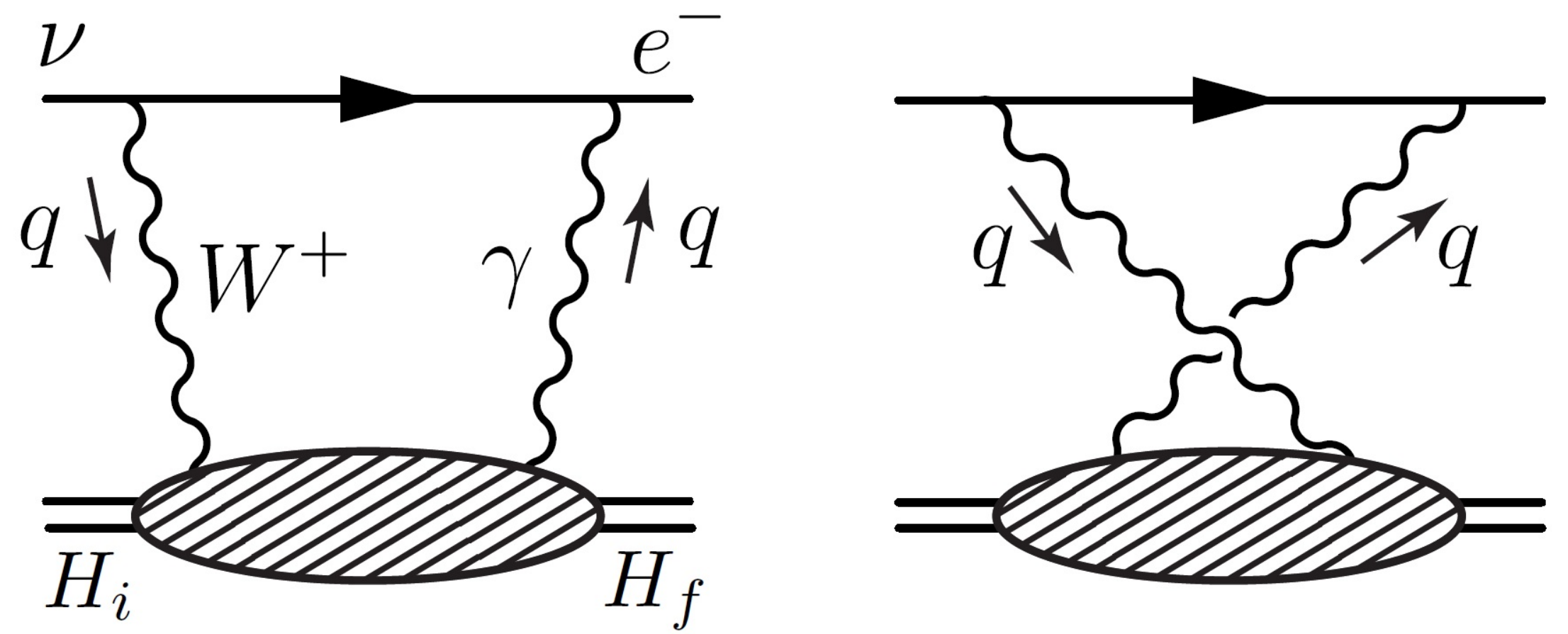}
         \caption{The $\gamma W$-box
         diagrams for the semileptonic decay process $H_i\to H_fe\bar{\nu}_e$.}
    \label{fig:photon_W_diags}
     \end{figure}

Sirlin's representation is particularly useful in the treatment of the
semileptonic decay $H_i\to H_f e\bar{\nu}_e$ with the hadrons $H_i$ and $H_f$ having
nearly the same masses $m_i\approx m_f$. In this case, the $O(G_F\alpha_e)$ RCs to the decay width is given
as~\cite{Sirlin:1977sv}
\be
\label{eq:delta}
\delta=\frac{\alpha_e}{2\pi}\left[\bar{g}+3\ln\frac{m_Z}{m_p}+\ln\frac{m_Z}{m_W}+\tilde{a}_g\right]+\delta_\mathrm{HO}^\mathrm{QED}+2\Box_{\gamma
W}^{VA},
\ee
where $m_Z$ and $m_W$ are the masses for the $Z$ and $W$ bosons. $m_p$ is the proton
mass that enters simply by convention. 
The Sirlin's function $\bar{g}$, which is a function of the electron's end-point energy, summarizes the  infrared-singular contributions involving both the one-loop and bremsstrahlung corrections~\cite{Sirlin:1967zza,Sirlin:1977sv,Wilkinson:1970cdv}.
The $O(\alpha_s)$ QCD correction $\tilde{a}_g$ is dominated by the high-energy
scale $Q^2\simeq m_W^2$ with a relatively small contribution of
$\frac{\alpha_e}{2\pi}\tilde{a}_g\approx -9.6\times10^{-5}$~\cite{Sirlin:1977sv,Seng:2019lxf}. The contribution
from the resummation of the large QED logs is contained in
$\delta_{\mathrm{HO}}^\mathrm{QED}=0.0010(3)$~\cite{Erler:2002mv}. All the
contributions that are sensitive
to hadronic scales, reside in the axial $\gamma W$-box contribution $\Box_{\gamma
W}^{VA}$, as shown in Fig.~\ref{fig:photon_W_diags}.
The total contribution $\delta$ is equivalent to $\left(S_{EW}-1\right)+\delta_K^{e}$ shown
in Eq.~(\ref{eq:master_formula}).

     \begin{figure}[htb]
     \centering
         \includegraphics[width=0.9\textwidth,angle=0]{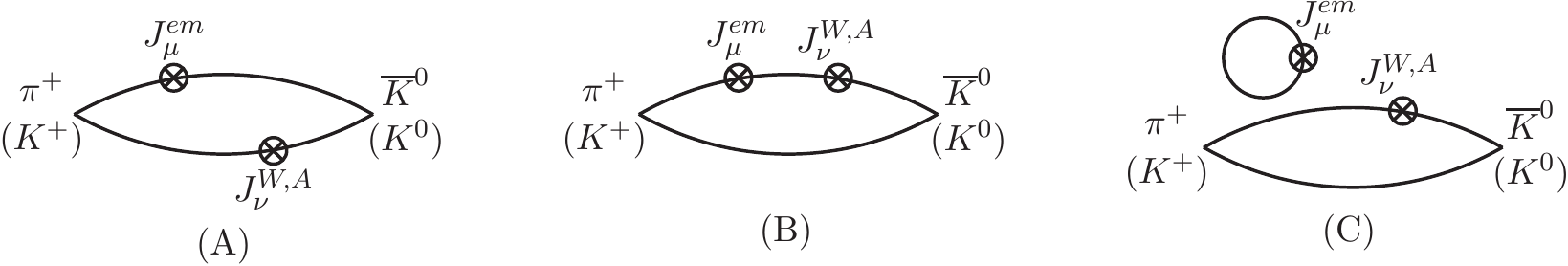}
         \caption{Quark contractions for $\overline{K}^0\to\pi^+e\bar{\nu}_e$ and
         $K^0\to K^+e\bar{\nu}_e$.}
    \label{fig:contraction}
     \end{figure}

In the $K_{\ell3}$ decays, since $m_K$ is not close to $m_\pi$, the non-perturbative hadronic
effects are contained not only in $\Box_{\gamma W}^{VA}$, but also in other
diagrams. As a consequence, Eq.~(\ref{eq:delta}) can not be used directly. To
evaluate the total RCs, the calculation of the five-point correlation function
is required. To simplify this problem, Ref.~\cite{Seng:2020jtz} proposes to calculate the RCs
for $\overline{K}^0\to\pi^+ e\bar{\nu}_e$ in the flavor SU(3) limit, where
$m_K=m_\pi$. The relevant contractions are shown in Fig.~\ref{fig:contraction} with
the disconnected diagram (C) vanishing in the flavor SU(3) limit. Although the
physical value of $\delta$ cannot be determined directly using this unphysical
setup, the lattice calculation can help to extract the LECs for ChPT. Then by
using ChPT one can obtain the physical RCs. Besides for the $\overline{K}^0\to\pi^+$
transition, the semileptonic decay of $K^0\to K^+ e\bar{\nu}_e$ can also be used
to determine the same LECs as it has the same contractions as
$\overline{K}^0\to\pi^+$ up to the disconnected parts.

In ChPT, the RCs to $K_{\ell3}$ are computed to
$O(e^2p^2)$~\cite{Cirigliano:2001mk,Cirigliano:2002ng,Cirigliano:2008wn} with the
short-distance radiative correction
\be
S_{EW}=1-e^2\left[-\frac{1}{2\pi^2}\ln\frac{M_Z}{M_\rho}+(X_6^{\mathrm{phys}})_{\alpha_s}\right]+\delta_{\mathrm{HO}}^{\mathrm{QED}}=1.0229(3),
\ee
where $M_\rho=0.77$ GeV is the rho mass and
$(X_6^{\mathrm{phys}})_{\alpha_s}\approx 3.0\times10^{-3}$~\cite{DescotesGenon:2005pw} summarizes the
$O(\alpha_s)$ pQCD contribution to $X_6^{\mathrm{phys}}$ with
$X_6^{\mathrm{phys}}(\mu)\equiv X_6^r(\mu)-4K_{12}^r(\mu)$ the combination of
two renormalized LECs. The scale $\mu$ is usually taken as $\mu=M_\rho$ in the numerical analysis.
The long-distance radiative correction $\delta_K^\ell$ has the dependence on the
LECs through the relation\footnote{Notice that in a similar expression in Ref.~\cite{Seng:2020jtz}, the quantity $\delta_{K^\pm}^\ell$ includes also contributions from the LECs $\{K_i^r\}$. That, however, was not the standard convention adopted by the ChPT community, which chooses to lump the $\{K_i^r\}$ contribution into $\delta_{SU2}$.}
\ba
\label{eq:LECs}
&&\delta_{K^\pm}^{\ell}=2e^2\left[-\frac{8}{3}X_1-\frac{1}{2}\tilde{X}_6^{\mathrm{phys}}(M_\rho)\right]+\cdots
\nn\\
&&\delta_{K^0}^{\ell}=2e^2\left[\frac{4}{3}X_1-\frac{1}{2}\tilde{X}_6^{\mathrm{phys}}(M_\rho)\right]+\cdots,
\ea
where the ellipses indicates the omission of the known kinematic terms,
which does not depend on the LECs. $X_1$ and $\tilde{X}_6^\mathrm{phys}$ are LECs relevant at $O(e^2p^2)$. $\tilde{X}_6^{\mathrm{phys}}(M_\rho)\equiv
X_6^{\mathrm{phys}}(M_\rho)+(2\pi^2)^{-1}\ln(M_Z/M_\rho)-(X_6^{\mathrm{phys}})_{\alpha_s}$
removes the large electroweak logarithm and the $O(\alpha_s)$ pQCD correction
from $X_6^{\mathrm{phys}}$.
In a similar way, one can define the quantity $\delta_{\pi^\pm}^{\ell}$ for
$\pi_{\ell3}$
\be
\delta_{\pi^{\pm}}^{\ell}=2e^2\left[-\frac{2}{3}X_1-\frac{1}{2}\tilde{X}_6^{\mathrm{phys}}(M_\rho)\right]+\cdots.
\ee

Since the neutral kaon decay mode $\overline{K}^0\to\pi^+e\bar{\nu}_e$ is
theoretically cleaner as it does not receive contributions from the
$\pi^0$-$\eta$ mixing which complicates the analysis in the flavor SU(3) limit,
we may use it to extract the LECs. 
Comparing the ChPT and Sirlin's representations,
the relation between the axial $\gamma W$-box contribution $\Box_{\gamma
W}^{VA}\big|_{K^0,\mathrm{SU(3)}}$ and the LECs is given by~\cite{Seng:2020jtz}
\be
\label{eq:box_to_LECs_kaon}
-\frac{8}{3}X_1+\bar{X}_6^{\mathrm{phys}}(M_\rho)=-\frac{1}{2\pi\alpha}
\left(\Box_{\gamma
W}^{VA}\big|_{K^0,\mathrm{SU(3)}}-\frac{\alpha}{8\pi}\ln\frac{M_W^2}{M_\rho^2}\right)+\frac{1}{8\pi^2}\left(\frac{5}{4}-\tilde{a}_g\right).
\ee
with $\bar{X}_6^{\mathrm{phys}}(M_\rho)$ defined as
$\bar{X}_6^{\mathrm{phys}}(M_\rho)\equiv
X_6^{\mathrm{phys}}(M_\rho)+(2\pi^2)^{-1}\ln(M_Z/M_\rho)$, which removes only the large electroweak logarithm but retains the full pQCD corrections.
For the $\pi_{\ell3}$ decay, the relation is given by
\be
\label{eq:box_to_LECs_pion}
\frac{4}{3}X_1+\bar{X}_6^{\mathrm{phys}}(M_\rho)=-\frac{1}{2\pi\alpha}
\left(\Box_{\gamma
W}^{VA}\big|_{\pi}-\frac{\alpha}{8\pi}\ln\frac{M_W^2}{M_\rho^2}\right)+\frac{1}{8\pi^2}\left(\frac{5}{4}-\tilde{a}_g\right).
\ee
The box contribution $\Box_{\gamma W}^{VA}\big|_{\pi}$ for the $\pi_{\ell3}$
decay has been calculated in Ref.~\cite{Feng:2020zdc}. The focus of this paper is 
on the determination of
$\Box_{\gamma W}^{VA}\big|_{K^0,\mathrm{SU(3)}}$, from which the
LECs $X_1$ and $\bar{X}_6^{\mathrm{phys}}(M_\rho)$ can be obtained.

The lattice QCD calculation of $\Box_{\gamma W}^{VA}\big|_{K^0,\mathrm{SU(3)}}$
can follow the procedures given in Ref.~\cite{Feng:2020zdc}.
We first define the hadronic function $\mathcal{H}^{VA}_{\mu\nu}(t,\vec{x})$ in
Euclidean space
\be
\mathcal{H}^{VA}_{\mu\nu}(t,\vec{x})\equiv
\langle
\pi^+(P)|T\left[J^{em}_\mu(t,\vec{x})J^{W,A}_\nu(0)\right]|\overline{K}^0(P)\rangle,
\ee
where $J^{em}_\mu=\frac{2}{3}\bar u \gamma_\mu u - \frac{1}{3} \bar d \gamma_\mu d-
\frac{1}{3} \bar s \gamma_\mu s$
is the electromagnetic quark current,
and $J^{W,A}_\nu=\bar u \gamma_\nu \gamma_5 s$ is the axial part
of the weak charged current.
The Euclidean momentum $P$ is chosen as
$P=(im_K,\vec{0})$ with $m_K=m_\pi$ in the flavor SU(3) limit.
The box contribution $\Box_{\gamma W}^{VA}\big|_{K^0,\mathrm{SU(3)}}$ can be
determined through the integral
\be
\label{eq:master}
\Box_{\gamma
W}^{VA}\big|_{K^0,\mathrm{SU(3)}}=\frac{3\alpha_e}{2\pi}\int\frac{dQ^2}{Q^2}\,\frac{m_W^2}{m_W^2+Q^2}M_{K}(Q^2)
\ee
with
\ba
\label{eq:lattice_M}
&&M_{K}(Q^2)=-\frac{1}{6}\frac{\sqrt{Q^2}}{m_K}\int
d^4x\,\omega(t,\vec{x})\epsilon_{\mu\nu\alpha0}x_\alpha\mathcal{H}^{VA}_{\mu\nu}(t,\vec{x}),
\nn\\
&&\omega(t,\vec{x})=\int_{-\frac{\pi}{2}}^{\frac{\pi}{2}}\frac{\cos^3\theta\,d\theta}{\pi}\frac{j_1\left(\sqrt{Q^2}|\vec{x}|\cos\theta\right)}{|\vec{x}|}
\cos\left(\sqrt{Q^2}t\sin\theta\right).
\ea
Here $j_1(x)$ is the spherical Bessel function. To compute the integral in
Eq.~(\ref{eq:master}), for small $Q^2$, we use lattice QCD input of
$\mathcal{H}^{VA}_{\mu\nu}(t,\vec{x})$.
For large $Q^2$, the operator product expansion of $J_{\mu}^{em}(x)J_{\nu}^{W,A}(0)$ is utilized with the Wilson
coefficients given at the four-loop accuracy~\cite{Larin:1991tj,Baikov:2010je}.
For more details, we refer the readers to Ref.~\cite{Feng:2020zdc}.

\section{Numerical results}

\begin{table}[htbp]
	\small
	\centering
	\begin{tabular}{cccccc}
		\hline
        \hline
        \multicolumn{2}{c}{Ensemble}  & $m_\pi$ [MeV] & $L$ &  $T$ & $a^{-1}$ [GeV] \\
        \hline
        & 24D  & 141.2(4) & $24$ & $64$ & $1.015$  \\
     DSDR   & 32D  & 141.4(3) & $32$ & $64$ & $1.015$ \\
        & 32D-fine & 143.0(3) & $32$ & $64$ & $1.378$ \\
        \hline
        \multirow{2}{*}{Iwasaki}   & 48I & 135.5(4) & $48$ & $96$ & $1.730$ \\
        & 64I & 135.3(2) & $64$ & $128$ & $2.359$ \\
        \hline
    \end{tabular}%
    \caption{Ensembles used in this work. For each ensemble we list the pion mass $m_\pi$, 
    the spatial and temporal extents, $L$ and $T$, 
    the inverse of lattice
    spacing $a^{-1}$.}
    \label{tab:ensemble_parameter}%
\end{table}%

Five gauge ensemble with $N_f=2+1$-flavor domain
wall fermion are used in the calculation. The detailed information is shown in Table~\ref{tab:ensemble_parameter}.
Here 48I and 64I
use the Iwasaki gauge action in the simulation (denoted as Iwasaki) while the
other three ensembles use Iwasaki+DSDR action (denoted as DSDR).
We place the Coulomb gauge-fixed wall-source quark propagators on all time slices. We calculate point-source propagators
at $O(1000)$ random spacetime locations. The correlation functions are
constructed
using the field sparsening technique~\cite{Li:2020hbj,Detmold:2019fbk} with a significant reduction
in the propagator storage. For the locations of two current insertions $J_\mu^{em}$ and
$J_\nu^{W,A}$, we treat one as the
source of the propagator and the other as the sink. In this way the hadronic
function $\mathcal{H}^{VA}_{\mu\nu}(t,\vec{x})$,
which depends
on the coordinate-space variable $x$
can be obtained. Such technique has also been used in the
computation of three-point correction function to extract the pion charge radius~\cite{Feng:2019geu}.
The flavor SU(3) limit is achieved by tuning down the strange quark mass to be
the same as the light quark mass.

     \begin{figure}[htb]
     \centering
         \includegraphics[width=0.9\textwidth,angle=0]{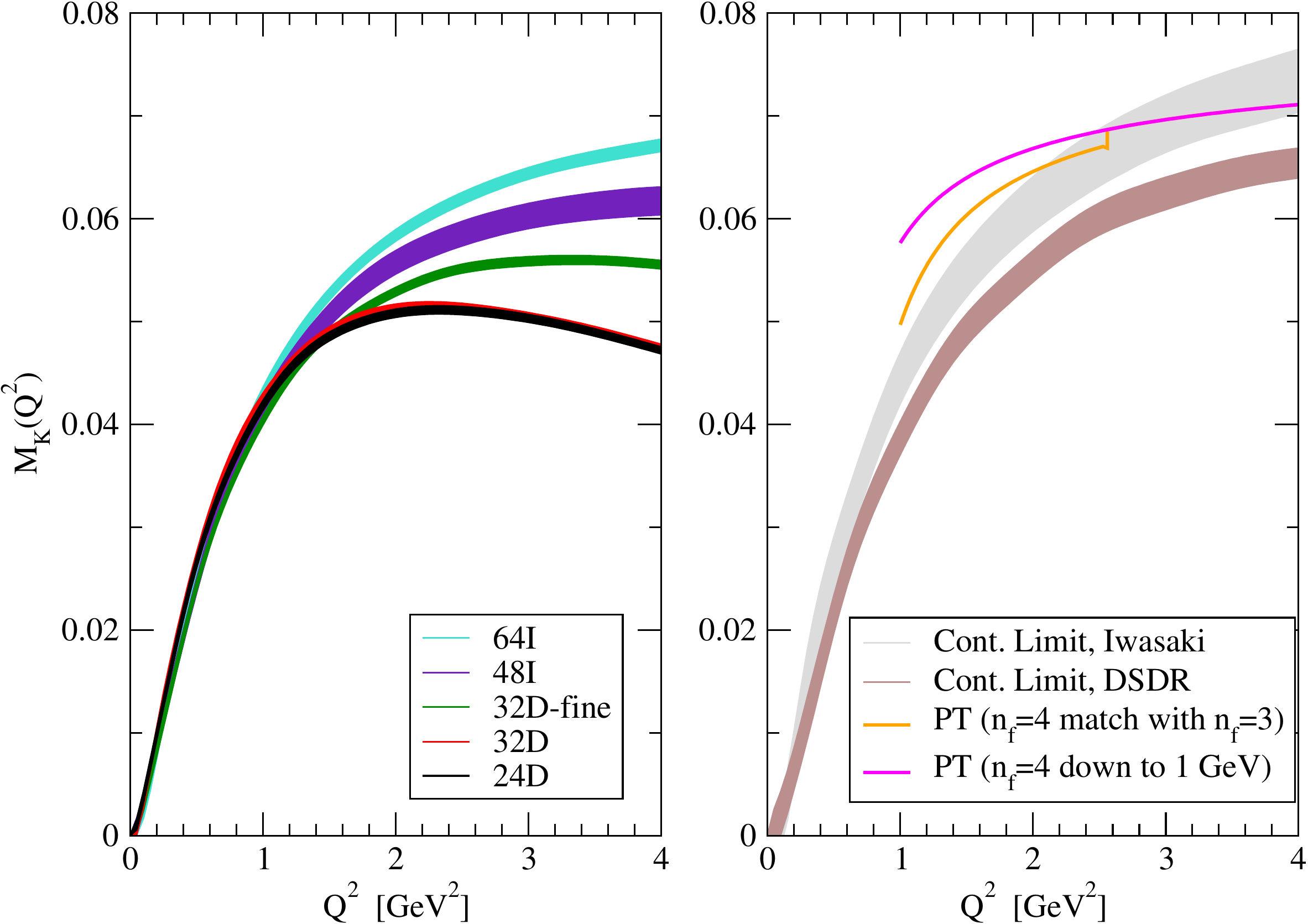}
         \caption{$M_K(Q^2)$ as a function of $Q^2$. In the left panel, 
         the lattice results for all five ensembles
         are given. 
         In the right panel, we have extrapolated the Iwasaki and DSDR results
         to their continuum limit. The remaining orange and magenta curves
         are the results from perturbation theory.}
         \label{fig:m_qsq}
    \end{figure}

Inserting $\mathcal{H}^{VA}_{\mu\nu}(t,\vec{x})$ into the
integral~(\ref{eq:lattice_M}), we calculate the scalar function $M_K(Q^2)$.
The lattice results for $M_K(Q^2)$ as a function of $Q^2$ are shown in
the left panel of Fig.~\ref{fig:m_qsq}.
At large $Q^2$ ($Q^2\gtrsim 1$ GeV$^2$), the lattice results from different
gauge ensembles start to disagree, suggesting the obvious
lattice discretization effects. In the right panel of Fig.~\ref{fig:m_qsq}, a
continuum extrapolation is performed to obtain the results in the continuum
limit for Iwasaki and DSDR ensembles separately. 
To reduce the systematic uncertainties contained in the lattice data at large $Q^2$, we
calculate the $M_K(Q^2)$ in pQCD using the RunDec
package~\cite{Chetyrkin:2000yt}.
At low $Q^2$ the perturbative results suffer from large pQCD truncation
effects
due to the lack of higher-loop and higher-twist contributions.
We observe an expected discrepancy between the orange and magenta curves at low $Q^2$, where 
the former uses the 4-flavor theory down to 1 GeV, while the latter 
turns to the 3-flavor theory upon decoupling the charm quark at 1.6 GeV.

\begin{table}[htbp]
	\small
	\centering
	\begin{tabular}{c|c|c|c}
		\hline
        \hline
        &
        \multicolumn{2}{c|}{$\Box_{\gamma W}^{VA,\le}\big|_{K^0,\mathrm{SU(3)}}$}
        & $\Box_{\gamma W}^{VA,>}\big|_{K^0,\mathrm{SU(3)}}$ \\
        \hline
        $Q^2_{\mathrm{cut}}$  & Iwasaki & DSDR & pQCD  \\
        \hline
        1 GeV$^2$ & $0.150(18)\times10^{-3}$ & $0.128(15)\times10^{-3}$ &
        $2.310(42)\times10^{-3}$ \\
        2 GeV$^2$ & $0.278(20)\times10^{-3}$ & $0.242(16)\times10^{-3}$ &
        $2.159(15)\times10^{-3}$ \\
        3 GeV$^2$ & $0.371(22)\times10^{-3}$ & $0.326(17)\times10^{-3}$ &
        $2.062(07)\times10^{-3}$ \\
        \hline
        \hline
    \end{tabular}%
    \caption{Using the scale $Q^2_{\mathrm{cut}}$ to split the integral range,
    the contributions of $\Box_{\gamma W}^{VA,\le}\big|_{K^0,\mathrm{SU(3)}}$ from lattice QCD and
    $\Box_{\gamma W}^{VA,>}\big|_{K^0,\mathrm{SU(3)}}$ from perturbation theory are shown. For the
    lattice results, we have performed the continuum extrapolation for Iwasaki
    and DSDR ensembles separately. For the perturbative results, the central
    values are compiled using the 4-flavor theory and uncertainties include 
    the higher-loop effects and the higher-twist effects with the error
    analysis following Ref.~\cite{Feng:2020zdc}.} 
    \label{tab:box_contribution}%
\end{table}%

We introduce a momentum-squared scale $Q^2_{\mathrm{cut}}$ that separates the
$Q^2$-integral into two regimes. We use the lattice data to determine the
integral for $Q^2\le Q^2_{\mathrm{cut}}$
and perturbation theory to determine the integral for $Q^2>Q^2_{\mathrm{cut}}$.
Three values of $Q^2_{\mathrm{cut}}=1,2,3$ GeV$^2$ are used to check the
$Q^2_{\mathrm{cut}}$-dependence in the final results. The corresponding
results for
$\Box_{\gamma W}^{VA,\le}\big|_{K^0,\mathrm{SU(3)}}$ and $\Box_{\gamma
W}^{VA,>}\big|_{K^0,\mathrm{SU(3)}}$ are listed
in Table~\ref{tab:box_contribution}. 

After combining the lattice data and perturbative results given in Table~\ref{tab:box_contribution}, we have
\be
\Box_{\gamma W}^{VA}\big|_{K^0,\mathrm{SU(3)}}=
\begin{cases}
    2.460(18)_{\mathrm{stat}}(42)_{\mathrm{PT}}(22)_{a}(1)_{\mathrm{FV}}\times10^{-3} & \mbox{$Q_{\mathrm{cut}}^2=1$ GeV$^2$} \\
    2.437(20)_{\mathrm{stat}}(15)_{\mathrm{PT}}(36)_{a}(1)_{\mathrm{FV}}\times10^{-3}
    & \mbox{$Q_{\mathrm{cut}}^2=2$ GeV$^2$}. \\
    2.433(22)_{\mathrm{stat}}(07)_{\mathrm{PT}}(45)_{a}(1)_{\mathrm{FV}}\times10^{-3} & \mbox{$Q_{\mathrm{cut}}^2=3$ GeV$^2$}
\end{cases}
\ee
Here we take the combination of the Iwasaki and perturbative results as the central value 
and estimate the residual lattice artifacts (with a subscript $a$) using the discrepancy between Iwasaki and DSDR.
The lattice finite-volume effects (with a subscript $\mathrm{FV}$) are estimated by comparing the
24D and 32D results. As a final result, we quote the value of $\Box_{\gamma
W}^{VA}\big|_{K^0,\mathrm{SU(3)}}$ at $Q_{\mathrm{cut}}^2=2$ GeV$^2$ and add the
statistical and systematic errors in quadrature
\be
\Box_{\gamma W}^{VA}\big|_{K^0,\mathrm{SU(3)}}=2.437(44)\times10^{-3}.
\ee

Inserting the result of $\Box_{\gamma W}^{VA}\big|_{K^0,\mathrm{SU(3)}}$ into
Eq.~(\ref{eq:box_to_LECs_kaon}), we obtain
\ba
\label{eq:LEC_kaon}
-\frac{8}{3}X_1+\bar{X}_6^{\mathrm{phys}}
=22.6(1.0)\times10^{-3}\quad\mbox{or}\quad
-\frac{8}{3}X_1+\tilde{X}_6^{\mathrm{phys}}=19.6(1.0)\times10^{-3}.
\ea
The previous ChPT analysis~\cite{Cirigliano:2008wn} quoted the LECs from the minimal resonance
model~\cite{Ananthanarayan:2004qk,DescotesGenon:2005pw} with
\be
X_1=-3.7(3.7)\times10^{-3},\quad
\tilde{X}_6^{\mathrm{phys}}=10.4(10.4)\times10^{-3}.
\ee
As it is hard to accurately estimate the uncertainty in these LECs
from the ChPT perspective, Ref.~\cite{Cirigliano:2008wn}
attributed to them a 100\%
uncertainty. Combining $X_1$ and $\tilde{X}_6^{\mathrm{phys}}$ together yields
\be
-\frac{8}{3}X_1+\tilde{X}_6^{\mathrm{phys}}=20.3(14.3)\times10^{-3}\quad\mbox{[minimal
resonance model]}.
\ee
Our result for $-\frac{8}{3}X_1+\tilde{X}_6^{\mathrm{phys}}$ agrees with
the minimal resonance model within few percent. Such a good 
agreement could easily be fortuitous as the methods used in the two studies
are very different and a large uncertainty is assigned to the estimate based on
the model.

For the $\pi_{\ell3}$ decay, substituting the lattice QCD result $\Box_{\gamma
	W}^{VA}\big|_{\pi}=2.830(28)\times10^{-3}$~\cite{Feng:2020zdc} into Eq.~(\ref{eq:box_to_LECs_pion})
yields
\be
\label{eq:LEC_pion}
\frac{4}{3}X_1+\bar{X}_6^{\mathrm{phys}}=14.0(6)\times 10^{-3}.
\ee
Combining Eqs.~(\ref{eq:LEC_kaon}) and (\ref{eq:LEC_pion}) together,
we have
\be
\label{eq:LECsnew}
X_1=-2.2(4)\times10^{-3},\quad
\bar{X}_6^{\mathrm{phys}}=16.9(7)\times10^{-3}.
\ee
Here the uncertainty is estimated conservatively through a linear addition.
It should be pointed out that in Eq.~(\ref{eq:LECsnew}) the estimate of the higher order
terms in the ChPT expansion are not included yet.

In ChPT, the RCs $\delta_{K}^\ell$ have two major sources of theoretical
uncertainties: the input of the LECs at $O(e^2p^2)$ and the unknown $O(e^2p^4)$
terms in the ChPT expansion. Using the LECs from this calculation, the former uncertainty
is significantly reduced, while the latter one remains. It results in an update of
$\delta_{K}^\ell$ (in units of \%)
\ba
&&\delta_{K^0}^e=0.99(19)_{e^2p^4}(11)_{\mathrm{LEC}}\quad\to\quad 1.00(19)
\nn\\
&&\delta_{K^0}^\mu=1.40(19)_{e^2p^4}(11)_{\mathrm{LEC}}\quad\to\quad
1.41(19)\nn\\
&&\delta_{K^{\pm}}^e=0.10(19)_{e^2p^4}(16)_{\mathrm{LEC}}\quad\to\quad -0.01(19)
\nn\\
&&\delta_{K^\pm}^\mu=0.02(19)_{e^2p^4}(16)_{\mathrm{LEC}}\quad\to\quad -0.09(19).
\ea
We refrain from presenting a corresponding update of $|V_{us}|$ in this paper,
because (1) our results for $\delta_K^\ell$ still agree with the existing
literature within error bars, and (2) our lattice calculation removes only the
LEC uncertainty but not the dominant, $\mathcal{O}(e^2p^4)$ uncertainty.
Therefore, we shall instead await a next round of global analysis in the near
future such as that in Ref.~\cite{Antonelli:2010yf}, whose impact on the precision
low-energy tests will be more significant. Our lattice result may serve as an important input to such an analysis.

\section{Conclusion}

Modern-day lattice QCD has reached the era when realistic calculations for many interesting second-order electroweak processes have become feasible.
In this work we perform a study of the
$\gamma W$-box correction to the kaon semileptonic decay $\overline{K}^0\to\pi^+
e\bar{\nu}_e$.
We adopt the new method proposed in Ref.~\cite{Seng:2020jtz}, which connects the Sirlin's
representation to the ChPT representation in the flavor SU(3) limit.
It allows us to determine the LECs for ChPT by computing the axial $\gamma
W$-box correction. We find that 
the values of the LECs devised from
the lattice calculation
agree well with the minimal resonance
model used in the literature, while a dramatic reduction of the respective uncertainties is achieved.
Finally, these LECs are used to estimate the RCs $\delta_{K}^{\ell}$ and help to reduce
its uncertainty. To further improve the determination of
RCs, the inclusion of higher-order terms in ChPT and the lattice QCD
computation of the complete set of Feynman diagrams are necessary.

\begin{acknowledgments}
X.F. and L.C.J. gratefully acknowledge many helpful discussions with our colleagues from the
RBC-UKQCD Collaborations.
X.F. and P.X.M. were supported in part by NSFC of China under Grants No.
    11775002 and No. 12070131001 and National Key Research and Development
    Program of China under Contracts No. 2020YFA0406400.
M.G.  is supported by EU Horizon 2020 research and innovation programme,
STRONG-2020 project, under grant agreement No 824093 and  by the German-Mexican
research collaboration Grant No. 278017 (CONACyT) and No. SP 778/4-1 (DFG).
L.C.J. acknowledges support by DOE Office of Science Early Career Award DE-SC0021147 and DOE grant DE-SC0010339.
The work of C.Y.S. is supported in part by the DFG
(Project-ID 196253076-TRR 110) and the NSFC (Grant No. 11621131001) through the funds
provided to the Sino-German CRC 110 ``Symmetries and the Emergence of Structure in QCD'',
and also by the Alexander von Humboldt Foundation through the Humboldt Research Fellowship.
The computation is performed under the ALCC Program of
the US DOE on the Blue Gene/Q (BG/Q) Mira computer at the Argonne Leadership Class Facility,
a DOE Office of Science Facility supported under Contract DE-AC02-06CH11357.
Computations for this work were carried out in part on facilities of the USQCD
Collaboration, which are funded by the Office of Science of the U.S. Department
of Energy.
The calculation is also carried out on Tianhe 3 prototype at Chinese National Supercomputer Center in Tianjin.
\end{acknowledgments}

\bibliography{paper}

\end{document}